\documentclass[12pt]{article}
\usepackage[margin=1.9cm]{geometry}
\usepackage{graphicx}
\usepackage{cite}
\usepackage{float}

\newcommand{\mysection}{\setcounter{equation}{0}\section}

\def\beq{\begin{equation}}
\def\eeq{\end{equation}}
\def\beqa{\begin{eqnarray}}
\def\eeqa{\end{eqnarray}}
 
\begin{document}

\begin{center}
{\Large \bf Soft-gluon corrections in $t{\bar t}W$ production}
\end{center}

\vspace{2mm}

\begin{center}
{\large Nikolaos Kidonakis and Chris Foster}\\

\vspace{2mm}

{\it Department of Physics, Kennesaw State University, \\
Kennesaw, GA 30144, USA}

\end{center}

\begin{abstract}
We study higher-order QCD corrections for the associated production of a top-antitop quark pair and a $W$ boson ($t{\bar t}W$ production) in proton-proton collisions. We calculate approximate NNLO (aNNLO) and approximate N$^3$LO (aN$^3$LO) cross sections, with second-order and third-order soft-gluon corrections added to the exact NLO QCD result, and we also include electroweak (EW) corrections through NLO. We calculate uncertainties from scale dependence, which are reduced at higher orders, and from parton distributions, and we also provide separate results for $t{\bar t}W^+$ and $t{\bar t}W^-$ production. We compare our results to recent measurements from the LHC, and we find that the aN$^3$LO QCD + NLO EW predictions provide improved agreement with the data. We also calculate differential distributions in top-quark transverse momentum and rapidity and find significant enhancements from the higher-order corrections. This is the first calculation of soft-gluon corrections in single-particle-inclusive kinematics for a $2 \to 3$ process where all final-state particles are massive.
\end{abstract}

\mysection{Introduction}

The cross sections for the production of a top-antitop quark pair in association with a $W$ boson, i.e. $t{\bar t}W$ production, have been measured at the LHC at 7 TeV \cite{CMS7}, 8 TeV \cite{CMS8a,ATLAS8,CMS8b}, and 13 TeV \cite{ATLAS13a,CMS13a,ATLAS13b,CMS13b,ATLAS13c} energies. The measurements have become increasingly precise and have been consistently above the theoretical predictions. Therefore, it is crucial to have higher-order results for this process.

The next-to-leading-order (NLO) QCD corrections for this process were calculated in \cite{BCE,CE} and, including parton showers in \cite{GKPT,MPT,CKR} and higher jet multiplicities in \cite{BRM}, while electroweak corrections were calculated in \cite{FHPSZ,FPZ} (see also \cite{DFSS}). Next-to-next-to-leading-order (NNLO) QCD corrections, though not exact for the finite part of the two-loop virtual corrections, were calculated in \cite{BDGKMRS}.

An important contribution to the higher-order corrections comes from the emission of soft (i.e. low energy) gluons. These contributions can be calculated at higher orders and formally resummed, and they are numerically dominant in all top-quark processes studied to date since the cross section receives large contributions from soft-gluon emission near partonic threshold because of the large top-quark mass. We use the soft-gluon resummation formalism of \cite{NKGS1,NKGS2,NK2loop} which has been applied to top-antitop pair production \cite{NKtt2l,NKttaN3LO,NKptyaN3LO,KGT} and single-top production \cite{NKsch,NKtW,NKtch} where it was shown that the soft-gluon corrections are very important. More recently, the resummation formalism has been extended \cite{FK2020} to $2 \to 3$ processes in single-particle-inclusive (1PI) kinematics, in particular $tqH$ production \cite{FK2021}, $tq\gamma$ production \cite{NKNY2022}, and $tqZ$ production \cite{NKNYtqZ}, as well as $t{\bar t} \gamma$ production \cite{NKAT}. In all these processes, the soft-gluon corrections are dominant and account for the majority of the complete corrections at NLO. Thus, it can reasonably be expected that they will also be important for $t{\bar t}W$ production. In this paper, we calculate cross sections from soft-gluon resummation using the formalism of \cite{NKtt2l,FK2020}. Alternative resummation formalisms for $t{\bar t}W$ production, with different kinematics, namely using the invariant mass of the $t{\bar t}W$ system, have been used in \cite{LLL,BFOP,KMSST,BFFPPT}. We note, however, that both the formalism and the choice of kinematics for the resummation can greatly affect the numerical values of the results (see e.g. the review paper of Ref. \cite{NKBP} for a discussion of different formalisms). Ours is the first calculation of soft-gluon corrections in 1PI kinematics for a $2 \to 3$ process where all final-state particles are massive.

We use the expansions of the resummed cross section to calculate approximate next-to-next-to-leading-order (aNNLO) and approximate next-to-next-to-next-to-leading-order (aN$^3$LO) cross sections and top-quark differential distributions for $t{\bar t}W$ production. The aNNLO results are derived by adding second-order soft-gluon corrections to the complete NLO calculation which includes QCD plus electroweak (EW) corrections. We find consistency between our aNNLO results and those in Ref. \cite{BDGKMRS} within uncertainties, which shows that the soft-gluon contributions are indeed dominant not only at NLO but also at NNLO. The aN$^3$LO results are derived by further adding third-order soft-gluon corrections to the aNNLO calculation.

In the next section, we describe the soft-gluon resummation formalism for $t{\bar t}W$ partonic processes. In Section 3, we present numerical results for the total cross section through aN$^3$LO QCD + NLO EW. In Section 4, we present results for the top-quark differential distributions in transverse momentum, $p_T$, and rapidity. We conclude in Section 5.   

\mysection{Resummation for $t{\bar t}W$ production}

We begin with the soft-gluon resummation formalism for $t{\bar t}W$ production. At leading order (LO), the parton-level processes are $a(p_a)+b(p_b) \to t(p_t)+{\bar t}(p_{\bar t})+W(p_W)$, with $a$ and $b$ denoting the two incoming partons (quarks and antiquarks), and we define the usual kinematical variables $s=(p_a+p_b)^2$, $t=(p_a-p_t)^2$, and $u=(p_b-p_t)^2$, as well as $s'=(p_t+p_{\bar t})^2$, $t'=(p_b-p_{\bar t})^2$, and $u'=(p_a-p_{\bar t})^2$. If an additional gluon is emitted in the final state, then momentum conservation gives $p_a +p_b=p_t +p_{\bar t} +p_W+p_g$ where $p_g$ is the gluon momentum. We denote the top-quark mass by $m_t$, and we define a threshold variable $s_4=(p_{\bar t}+p_W+p_g)^2-(p_{\bar t}+p_W)^2=s+t+u-m_t^2-(p_{\bar t}+p_W)^2$ which involves the extra energy from gluon emission and which vanishes as $p_g \to 0$. In our 1PI kinematics, with the top quark being the observed particle, the soft-gluon corrections appear in the perturbative series as terms with coefficients multiplying ``plus distributions'' of logarithms of $s_4$, specifically $\ln^k(s_4/m_t^2)/s_4$ where $k$ takes integer values from 0 through $2n-1$ for the $n$th order corrections.

Resummation follows from factorization properties of the cross section and renormalization-group evolution \cite{NKGS1,NKGS2,FK2020,LOS}. We begin by writing the differential cross section for $t{\bar t}W$ production in proton-proton collisions as a convolution, 
\beq
d\sigma_{pp \to t{\bar t}W}=\sum_{a,b} \; 
\int dx_a \, dx_b \,  \phi_{a/p}(x_a, \mu_F) \, \phi_{b/p}(x_b, \mu_F) \, 
d{\hat \sigma}_{ab \to t{\bar t}W}(s_4, \mu_F) \, ,
\label{sigma}
\eeq
where $\mu_F$ is the factorization scale, $\phi_{a/p}$  and $\phi_{b/p}$ are parton distribution functions (pdf) for parton $a$ and parton $b$, respectively, in the proton, and ${\hat \sigma}_{ab \to t{\bar t}W}$ is the partonic cross section which at a given order also depends on the renormalization scale $\mu_R$.

The cross section factorizes if we take Laplace transforms, defined by 
\beq
 {\tilde{\hat\sigma}}_{ab \to t{\bar t}W}(N)=\int_0^{s_4^{\rm max}} \frac{ds_4}{s} \,  e^{-N s_4/s} \; {\hat\sigma}_{ab \to t{\bar t}W}(s_4), 
\eeq
where $N$ is the transform variable. Under transforms, the logarithms of $s_4$ in the perturbative series turn into logarithms of $N$ which, as we will see, exponentiate. We also define transforms of the pdf via ${\tilde \phi}(N)=\int_0^1 e^{-N(1-x)} \phi(x) \, dx$. Replacing the colliding protons by partons in Eq. (\ref{sigma}) \cite{NKGS2,FK2020,GS}, we thus have a factorized form in transform space
\beq
d{\tilde \sigma}_{ab \to t{\bar t}W}(N)= {\tilde \phi}_{a/a}(N_a, \mu_F) \, {\tilde \phi}_{b/b}(N_b, \mu_F) \, d{\tilde{\hat \sigma}}_{ab \to t{\bar t}W}(N, \mu_F) \, .
\label{factcs}
\eeq

The cross section can be refactorized \cite{NKGS1,NKGS2,FK2020} in terms of an infrared-safe short-distance hard function, $H_{ab \to t{\bar t}W}$,  and a soft function, $S_{ab \to t{\bar t}W}$, which describes the emission of noncollinear soft gluons. These hard and soft functions are $2\times 2$ matrices in the color space of the partonic scattering. We have
\beq
d{\tilde{\sigma}}_{ab \to t{\bar t}W}(N)={\tilde \psi}_{a/a}(N_a,\mu_F) \, {\tilde \psi}_{b/b}(N_b,\mu_F) \, {\rm tr} \left\{H_{ab \to t{\bar t}W} \left(\alpha_s(\mu_R)\right) \, {\tilde S}_{ab \to t{\bar t}W} \left(\frac{\sqrt{s}}{N \mu_F} \right)\right\} \, ,
\label{refactcs}
\eeq
where the functions $\psi$ are distributions for incoming partons at fixed value of momentum and involve collinear emission \cite{NKGS1,NKGS2,FK2020,GS}.

Comparing Eqs. (\ref{factcs}) and (\ref{refactcs}), we find an expression for the hard-scattering partonic cross section in transform space
\beq
d{\tilde{\hat \sigma}}_{ab \to t{\bar t}W}(N, \mu_F)=
\frac{{\tilde \psi}_{a/a}(N_a, \mu_F) \, {\tilde \psi}_{b/b}(N_b, \mu_F)}{{\tilde \phi}_{a/a}(N_a, \mu_F) \, {\tilde \phi}_{b/b}(N_b, \mu_F)} \; \,  {\rm tr} \left\{H_{ab \to t{\bar t}W}\left(\alpha_s(\mu_R)\right) \, {\tilde S}_{ab \to t{\bar t}W}\left(\frac{\sqrt{s}}{N \mu_F} \right)\right\} \, .
\label{sigN}
\eeq

The dependence of the soft matrix on the transform variable, $N$, is resummed via renormalization-group evolution \cite{NKGS1,NKGS2}. Thus, ${\tilde S}_{ab \to t{\bar t}W}$ obeys a renormalization-group equation involving a soft anomalous dimension matrix, $\Gamma_{\! S \, ab \to t{\bar t}W}$, which is calculated from the coefficients of the ultraviolet poles of the eikonal diagrams for the partonic processes \cite{NKGS1,NKGS2,NK2loop,NKtt2l,FK2020}.

The $N$-space resummed cross section, which resums logarithms of $N$, is derived from the renormalization-group evolution of the functions ${\tilde S}_{ab \to t{\bar t}W}$, ${\tilde \psi}_{a/a}$, ${\tilde \psi}_{b/b}$, ${\tilde \phi}_{a/a}$, and ${\tilde \phi}_{b/b}$ in Eq. (\ref{sigN}), and it is given by
\beqa
d{\tilde{\hat \sigma}}_{ab \to t{\bar t}W}^{\rm resum}(N,\mu_F) &=&
\exp\left[\sum_{i=a,b} E_{i}(N_i)\right] \, 
\exp\left[\sum_{i=a,b} 2 \int_{\mu_F}^{\sqrt{s}} \frac{d\mu}{\mu} \gamma_{i/i}(N_i)\right]
\nonumber\\ && \hspace{-5mm} \times \,
{\rm tr} \left\{H_{ab \to t{\bar t}W}\left(\alpha_s(\sqrt{s})\right) {\bar P} \exp \left[\int_{\sqrt{s}}^{{\sqrt{s}}/N}
\frac{d\mu}{\mu} \; \Gamma_{\! S \, ab \to t{\bar t}W}^{\dagger} \left(\alpha_s(\mu)\right)\right] \; \right.
\nonumber\\ && \left. \hspace{5mm} \times \,
{\tilde S}_{ab \to t{\bar t}W} \left(\alpha_s\left(\frac{\sqrt{s}}{N}\right)\right) \;
P \exp \left[\int_{\sqrt{s}}^{{\sqrt{s}}/N}
\frac{d\mu}{\mu}\; \Gamma_{\! S \, ab \to t{\bar t}W}
\left(\alpha_s(\mu)\right)\right] \right\} \, ,
\label{resummed}
\eeqa
where $P$ (${\bar P}$) denotes path-ordering in the same (reverse) sense as the integration variable $\mu$.
The first exponential in Eq. (\ref{resummed}) resums soft and collinear emission from the initial-state partons \cite{GS}, while the second exponential involves the parton anomalous dimensions $\gamma_{i/i}$ and the factorization scale $\mu_F$. These are followed by exponentials with the soft anomalous dimension matrix $\Gamma_{\! S \, ab \to t{\bar t}W}$ and its Hermitian adjoint. Explicit results for the first two exponents of Eq. (\ref{resummed}) can be found in Ref. \cite{FK2020}. The trace over the product of the LO hard and soft matrices provides the LO cross section, while the NLO cross section, to which we match, involves the NLO hard and soft matrices. 

The soft anomalous dimensions for $t{\bar t} W$ production are essentially the same as those for the $q{\bar q}$ channel in $t{\bar t}$ production \cite{NKGS1,NKGS2,NKtt2l,FNPY}, since the color structure of the hard scattering is the same, with only some modifications to account for the $2 \to 3$ kinematics (see Ref. \cite{NKuni} for a review). For the processes  $q(p_a)+{\bar q}'(p_b) \to t(p_t)+{\bar t}(p_{\bar t})+W(p_W)$ we choose a color tensor basis of $s$-channel singlet and octet exchange, $c_1^{q{\bar q}' \rightarrow t{\bar t}W} = \delta_{ab}\delta_{12}$,  $c_2^{q{\bar q}' \rightarrow t{\bar t}W} =  T^c_{ba} \, T^c_{12}$, where $T^c$ are the generators of SU(3) in the fundamental representation with color indices 1 and 2 for the top and antitop quarks, respectively. Then, the four matrix elements of $\Gamma_{\!\! S \, q{\bar q}'\rightarrow t{\bar t}W}$ are given at one loop \cite{NKuni} by 
\beqa
&& \hspace{-7mm} \Gamma_{\!\! 11 \, q{\bar q}' \to t{\bar t}W}^{(1)} = \Gamma_{\rm cusp}^{(1)},
\quad
\Gamma_{\!\! 12 \, q{\bar q}' \to t{\bar t}W}^{(1)}=
\frac{C_F}{2N_c} \Gamma_{\!\! 21 \, q{\bar q}' \to t{\bar t}W}^{(1)} ,
\quad
\Gamma_{\!\! 21 \, q{\bar q}' \to t{\bar t}W}^{(1)}=
\ln\left(\frac{(t-m_t^2)(t'-m_t^2)}{(u-m_t^2)(u'-m_t^2)}\right),
\nonumber \\ && \hspace{-7mm}
\Gamma_{\!\! 22 \, q{\bar q}' \to t{\bar t}W}^{(1)} = \left(1-\frac{C_A}{2C_F}\right)
\left[\Gamma_{\rm cusp}^{(1)}+2C_F\ln\left(\frac{(t-m_t^2)(t'-m_t^2)}{(u-m_t^2)(u'-m_t^2)}\right)\right]+\frac{C_A}{2}\left[\ln\left(\frac{(t-m_t^2)(t'-m_t^2)}{s\, m_t^2}\right)-1\right]
\nonumber \\
\eeqa
where $\Gamma_{\rm cusp}^{(1)}=-C_F\left(L_{\beta_t}+1\right)$ is the one-loop QCD massive cusp anomalous dimension, with $L_{\beta_t}=(1+\beta_t^2)/(2\beta_t) \ln[(1-\beta_t)/(1+\beta_t)]$ and $\beta_t=\sqrt{1-4m_t^2/s'}$. Also, $C_F=(N_c^2-1)/(2N_c)$ and $C_A=N_c$, where $N_c=3$ in QCD.
 
At two loops, we have
\beqa
&& \hspace{-7mm} \Gamma_{\!\! 11 \, q{\bar q}' \to t{\bar t}W}^{(2)}=\Gamma_{\rm cusp}^{(2)}, \; \;
\Gamma_{\!\! 12 \, q{\bar q}' \to t{\bar t}W}^{(2)}=
\left(K_2-C_A \, N_2^{\beta_t}\right) \Gamma_{\!\! 12 \, q{\bar q}' \to t{\bar t}W}^{(1)}, \; \; 
\Gamma_{\!\! 21 \, q{\bar q}' \to t{\bar t}W}^{(2)}=
\left(K_2+C_A \, N_2^{\beta_t}\right) \Gamma_{\!\! 21 \, q{\bar q}' \to t{\bar t}W}^{(1)} ,
\nonumber \\ && \hspace{-7mm} 
\Gamma_{\!\! 22 \, q{\bar q}' \to t{\bar t}W}^{(2)}=
K_2 \, \Gamma_{\!\! 22 \, q{\bar q}' \to t{\bar t}W}^{(1)}
+\left(1-\frac{C_A}{2C_F}\right)
\left(\Gamma_{\rm cusp}^{(2)}-K_2 \, \Gamma_{\rm cusp}^{(1)}\right)
+\frac{1}{4} C_A^2(1-\zeta_3) \, ,
\eeqa
where
$K_2=C_A(67/36-\zeta_2/2)-5n_f/18$ with $n_f$ the number of light-quark flavors,
\beq
N_2^{\beta_t}=\frac{1}{4}\ln^2\left(\frac{1-\beta_t}{1+\beta_t}\right)
+\frac{(1+\beta_t^2)}{8 \beta_t} \left[\zeta_2
-\ln^2\left(\frac{1-\beta_t}{1+\beta_t}\right)
-{\rm Li}_2\left(\frac{4\beta_t}{(1+\beta_t)^2}\right)\right] \, ,
\eeq
and 
\beqa
\Gamma_{\rm cusp}^{(2)}&=& K_2 \, \Gamma_{\rm cusp}^{(1)}
+C_F C_A \left\{\frac{1}{2}+\frac{\zeta_2}{2}
+\frac{1}{2}\ln^2\left(\frac{1-\beta_t}{1+\beta_t}\right) \right.
\nonumber \\ &&  \hspace{-3mm}
{}+\frac{(1+\beta_t^2)}{4\beta_t}\left[\zeta_2\ln\left(\frac{1-\beta_t}{1+\beta_t}\right)-\ln^2\left(\frac{1-\beta_t}{1+\beta_t}\right)
+\frac{1}{3}\ln^3\left(\frac{1-\beta_t}{1+\beta_t}\right)
-{\rm Li}_2\left(\frac{4\beta_t}{(1+\beta_t)^2}\right)\right] 
\nonumber \\ &&  \hspace{-3mm} 
{}+\frac{(1+\beta_t^2)^2}{8\beta_t^2}\left[-\zeta_3-\zeta_2\ln\left(\frac{1-\beta_t}{1+\beta_t}\right)-\frac{1}{3}\ln^3\left(\frac{1-\beta_t}{1+\beta_t}\right)
-\ln\left(\frac{1-\beta_t}{1+\beta_t}\right) {\rm Li}_2\left(\frac{(1-\beta_t)^2}{(1+\beta_t)^2}\right) \right.
\nonumber \\ &&  \hspace{22mm} \left. \left.
{}+{\rm Li}_3\left(\frac{(1-\beta_t)^2}{(1+\beta_t)^2}\right)\right]\right\}
\eeqa
is the two-loop massive cusp anomalous dimension in QCD \cite{NK2loop,NK4loop}.

We can expand the resummed cross section, Eq. (\ref{resummed}), to any fixed order, and then do a straightforward inversion back to momentum space without requiring a prescription \cite{NKtt2l,NKttaN3LO,NKptyaN3LO,FK2020}. Thus, by expanding to N$^3$LO, we can calculate the first-order, second-order, and third-order soft-gluon corrections to the $t{\bar t}W$ production total cross section as well as the top-quark differential distributions. By adding the second-order soft-gluon corrections to the exact NLO results we derive aNNLO predictions, and by further adding the third-order soft-gluon corrections to the aNNLO results we derive aN$^3$LO predictions. 

We want to highlight the difference between our approach and those in earlier studies \cite{LLL,BFOP,KMSST,BFFPPT}. First of all, in general, when discussing formal logarithmic accuracy in resummation (i.e. NNLL) it is first important to specify what variable is used in these logarithms. In our 1PI approach we use logarithms of $s_4/m_t^2$ while previous work has used a variable involving the invariant mass of the $t{\bar t}W$ system. As is already well known from $t{\bar t}$ production, this choice has a big effect. The kinematics choice also affects the choice of central scale. In 1PI kinematics, it is natural to choose the mass of the observed top quark, $m_t$, as the central scale. Second, the formalism used, e.g. resummation in Mellin-moment space or under Laplace transforms or via SCET produces different results (see e.g. the discussion in \cite{NKBP}). Third, it is important to specify if any prescriptions are used (and their ambiguities) or if fixed-order expansions are employed. Thus, two calculations that have the same formal logarithmic accuracy may produce different numerical results and, in fact, this is the case among the various studies for $t {\bar t} W$ production as well as other processes. A litmus test of a formalism, however, is the success of the calculations in predicting later exact NNLO calculations. The formalism that we use here was the most successful \cite{NKtt2l} in predicting the NNLO corrections for $t{\bar t}$ production and, as we will discuss in the next section, it is in excellent agreement with the recent NNLO (partial) calculation in Ref. \cite{BDGKMRS}.

\mysection{Total cross sections}

In this section we present results for total cross sections for $t{\bar t}W$ production. We use a top-quark mass $m_t=172.5$ GeV and set the factorization and renormalization scales equal to each other, with this common scale denoted by $\mu$. As mentioned above, since we use 1PI kinematics, we make the natural choice of using the mass of the observed top quark, $m_t$, as the central scale in all our results below. The complete NLO results, which include both QCD and EW corrections, are calculated using {\small \sc MadGraph5\_aMC@NLO} \cite{MG5,MGew}. We add second-order soft-gluon corrections to the NLO QCD and NLO QCD+EW results to derive aNNLO QCD and aNNLO QCD + NLO EW cross sections, respectively. We further add third-order soft-gluon corrections to the aNNLO QCD and aNNLO QCD + NLO EW results to derive, respectively, aN$^3$LO QCD and aN$^3$LO QCD + NLO EW cross sections. We use MSHT20 NNLO pdf \cite{MSHT20} in our calculations.

\begin{figure}[htbp]
\begin{center}
\includegraphics[width=88mm]{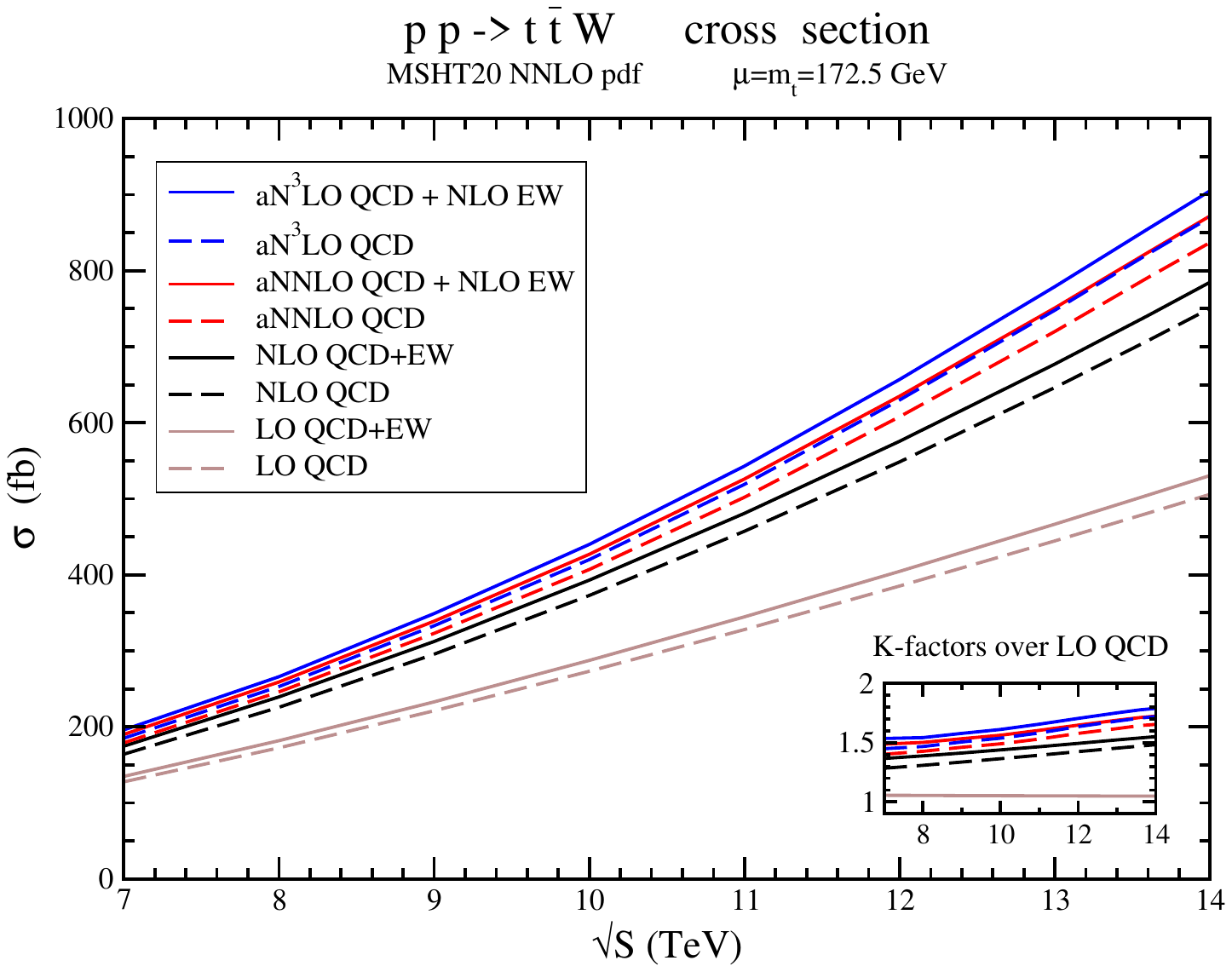}
\includegraphics[width=88mm]{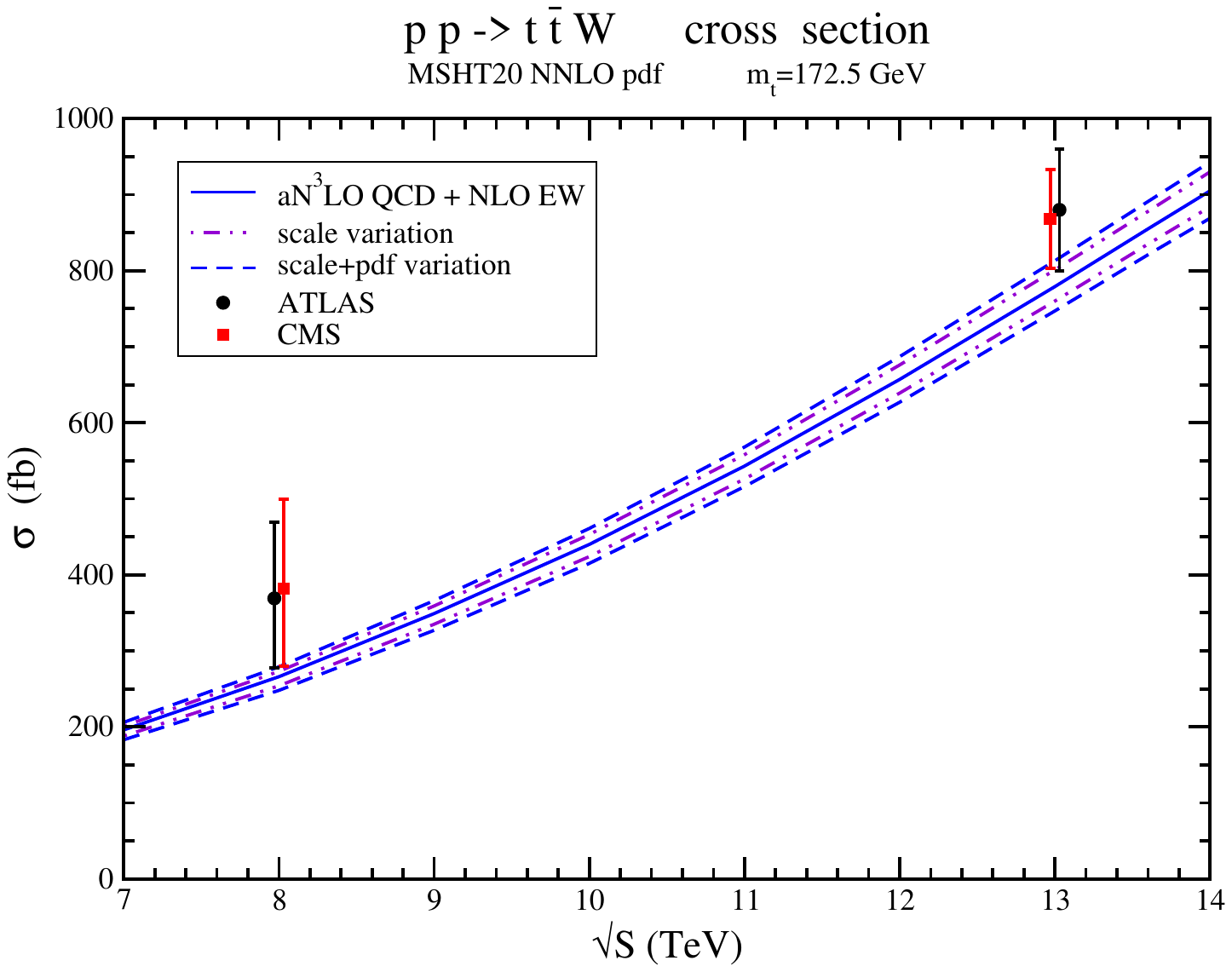}
\caption{The total cross sections at various orders through aN$^3$LO QCD + NLO EW for $t{\bar t}W$ production in $pp$ collisions at LHC energies using MSHT20 NNLO pdf. The inset in the left plot displays the $K$-factors relative to LO QCD while the plot on the right shows a comparison with LHC data at 8 TeV \cite{ATLAS8,CMS8b} and 13 TeV \cite{CMS13b,ATLAS13c} energies.}
\label{xsec}
\end{center}
\end{figure}

In Fig. \ref{xsec} we display the $t{\bar t}W$ cross sections in proton-proton collisions for LHC energies from 7 to 14 TeV. The plot on the left shows results at LO QCD, LO QCD+EW, NLO QCD, NLO QCD+EW, aNNLO QCD, aNNLO QCD + NLO EW, aN$^3$LO QCD, and aN$^3$LO QCD + NLO EW. The inset plot displays the $K$-factors, i.e. the ratios of the higher-order results to the LO QCD cross section. The results do not change materially if we use instead the MSHT20 aN$^3$LO pdf \cite{MSHT20aN3LO} or other pdf sets. It is clear that the NLO QCD corrections are large and the further enhancements from soft-gluon emission at aNNLO and aN$^3$LO are quite significant. The NLO electroweak corrections are also significant.

The plot on the right in Fig. \ref{xsec} compares the aN$^3$LO QCD + NLO EW theoretical prediction with data from ATLAS and CMS at 8 and 13 TeV energies \cite{CMS8b,ATLAS8,ATLAS13c,CMS13b}. In addition to the central result for the theoretical prediction, we also show the scale variation over the interval from $m_t/2$ to $2m_t$ as well as the combined scale and pdf variation. The theoretical results agree with the data within the theoretical and experimental uncertainties.

\begin{table}[H]
\begin{center}
\begin{tabular}{|c|c|c|c|c|c|c|c|c|} \hline
\multicolumn{6}{|c|}{$t{\bar t} W$ cross sections in $pp$ collisions at the LHC} \\ \hline
$\sigma$ in fb & 7 TeV & 8 TeV & 13 TeV & 13.6 TeV & 14 TeV \\ \hline
LO QCD              & $128^{+39}_{-28}$ & $172^{+51}_{-36}$ & $445^{+114}_{-84}$ & $481^{+121}_{-90}$ & $506^{+126}_{-94}$ \\ \hline
LO QCD+EW          & $135^{+41}_{-29}$ & $182^{+53}_{-38}$ & $467^{+119}_{-88}$ & $505^{+127}_{-94}$ & $531^{+132}_{-98}$ \\ \hline
NLO QCD             & $164^{+13}_{-17}$ & $226^{+20}_{-23}$ & $646^{+83}_{-74} $ & $708^{+94}_{-82}$ & $750^{+101}_{-88}$ \\ \hline
NLO QCD+EW          & $175^{+12}_{-17}$ & $239^{+19}_{-23}$ & $677^{+80}_{-74} $ & $741^{+90}_{-82}$ & $785^{+97}_{-88}$ \\ \hline
aNNLO QCD           & $179^{+6}_{-10}$ & $246^{+9}_{-15}$ & $720^{+29}_{-43} $ & $791^{+32}_{-47}$ & $837^{+34}_{-50}$ \\ \hline
aNNLO QCD + NLO EW  & $190^{+6}_{-10}$ & $259^{+9}_{-15}$ & $751^{+27}_{-43} $ & $824^{+29}_{-47}$ & $872^{+31}_{-50}$ \\ \hline
aN$^3$LO QCD         & $185^{+5}_{-8}$ & $253^{+7}_{-12}$ & $748^{+24}_{-19} $ & $822^{+26}_{-20}$ & $870^{+28}_{-21}$ \\ \hline
aN$^3$LO QCD + NLO EW & $196^{+5}_{-8}$ & $266^{+7}_{-12}$ & $779^{+22}_{-19} $ & $855^{+23}_{-20}$ & $905^{+25}_{-21}$ \\ \hline
\end{tabular}
\caption[]{The $t{\bar t}W$ cross sections (in fb) at various orders through aN$^3$LO QCD + NLO EW with scale uncertainties, in $pp$ collisions with $\sqrt{S}=7$, 8, 13, 13.6, and 14 TeV, $m_t=172.5$ GeV, and MSHT20 NNLO pdf.}
\label{table1}
\end{center}
\end{table}

In Table 1 we show detailed results for the LO QCD, LO QCD+EW, NLO QCD, NLO QCD+EW, aNNLO QCD, aNNLO QCD + NLO EW, aN$^3$LO QCD, and aN$^3$LO QCD + NLO EW cross sections for $t{\bar t}W$ production in $pp$ collisions at LHC energies of 7, 8, 13, 13.6, and 14 TeV. The central results are with $\mu=m_t=172.5$ GeV, and the scale uncertainty from variation over the interval $m_t/2$ to $2m_t$ is also shown. We checked that the scale uncertainty is not increased if we do a 7-point scale variation where $\mu_F$ and $\mu_R$ are varied independently. As expected, the scale uncertainty decreases with increasing perturbative order. At the current LHC energy of 13.6 TeV, the NLO QCD corrections increase the LO QCD result by around 47\%, the aNNLO corrections provide another 17\% increase, and the aN$^3$LO corrections provide an extra 6\% increase. The contribution of the electroweak corrections through NLO is around 7\%, similar in size to the aN$^3$LO QCD corrections. The final aN$^3$LO QCD + NLO EW cross section is around 78\% bigger than the LO QCD result. 

In Table 2 we show corresponding results separately for the cross sections for $t{\bar t}W^+$ and $t{\bar t}W^-$ production in $pp$ collisions at LHC energies of 13 and 13.6 TeV. The central results are with $\mu=m_t=172.5$ GeV, and the scale uncertainty from variation over the interval $m_t/2$ to $2m_t$ is also shown. Again, the scale uncertainties decrease with increasing perturbative order. The $t{\bar t}W^+$ cross sections are roughly twice as big as those for $t{\bar t}W^-$. We also find that the higher-order corrections are slightly bigger for the $t{\bar t}W^-$ process than for $t{\bar t}W^+$. 

\begin{table}[H]
\begin{center}
\begin{tabular}{|c|c|c|c|c|c|c|c|c|} \hline
\multicolumn{5}{|c|}{$t{\bar t} W^{+}$ and $t{\bar t} W^{-}$ cross sections in $pp$ collisions at the LHC} \\ \hline
$\sigma$ in fb & $t{\bar t} W^{+}$ 13 TeV & $t{\bar t} W^{+}$ 13.6 TeV & $t{\bar t} W^{-}$ 13 TeV & $t{\bar t} W^{-}$ 13.6 TeV \\ \hline
LO QCD              & $299^{+77}_{-57}$ & $322^{+82}_{-60}$ & $146^{+37}_{-28}$ & $159^{+40}_{-30}$ \\ \hline
LO QCD+EW           & $313^{+80}_{-59}$ & $337^{+85}_{-63}$ & $154^{+39}_{-29}$ & $168^{+42}_{-31}$ \\ \hline
NLO QCD             & $431^{+54}_{-49}$ & $470^{+61}_{-54}$ & $215^{+29}_{-25}$ & $238^{+33}_{-28}$ \\ \hline
NLO QCD+EW          & $450^{+51}_{-48}$ & $490^{+58}_{-53}$ & $227^{+28}_{-25}$ & $251^{+32}_{-28}$ \\ \hline
aNNLO QCD           & $480^{+19}_{-28}$ & $525^{+21}_{-31}$ & $240^{+10}_{-15}$ & $266^{+11}_{-16}$ \\ \hline
aNNLO QCD + NLO EW  & $499^{+17}_{-28}$ & $545^{+19}_{-31}$ & $252^{+10}_{-15}$ & $279^{+10}_{-16}$ \\ \hline
aN$^3$LO QCD        & $498^{+16}_{-12}$ & $545^{+17}_{-13}$ & $250^{+8}_{-7} $  & $277^{+9}_{-7}$ \\ \hline     
aN$^3$LO QCD + NLO EW & $517^{+14}_{-12}$ & $565^{+15}_{-13}$ & $262^{+8}_{-7}$ & $290^{+8}_{-7}$ \\ \hline
\end{tabular}
\caption[]{The separate $t{\bar t}W^{+}$ and $t{\bar t}W^{-}$ cross sections (in fb) at various orders through aN$^3$LO QCD + NLO EW with scale uncertainties, in $pp$ collisions with $\sqrt{S}=13$ and 13.6 TeV, $m_t=172.5$ GeV, and MSHT20 NNLO pdf.}
\label{table2}
\end{center}
\end{table}

We also want to note that our results are consistent with those in \cite{BDGKMRS} through NNLO QCD + NLO EW, within the uncertainties provided, if we use their choices for the central scale and pdf. We reproduce the LO QCD, NLO QCD, and NLO EW corrections and, more importantly, we find that at NNLO QCD we find excellent agreement. More specifically, at 13 TeV, the NNLO QCD result for the $t{\bar t}W$ cross section in Table I of \cite{BDGKMRS} is $711^{+35}_{-46} \pm 14$ fb where the first uncertainty is from scale variation and the second from the fact that the calculation is not exact for the finite part of the two-loop virtual corrections. Using the same parameters and scale and pdf choices as \cite{BDGKMRS}, we find $708^{+31}_{-45}$ fb where the uncertainty is from the seven-point variation around a central scale $\mu=m_t+m_W/2$ as used in \cite{BDGKMRS}. Thus, the two results are basically the same in both central value and scale uncertainty. This again shows that the soft-gluon contributions are dominant in the higher-order corrections. The separate results for $t{\bar t}W^+$ and $t{\bar t}W^-$ in \cite{BDGKMRS} are, respectively, $475^{+23}_{-30} \pm 9$ fb and $236^{+12}_{-16} \pm 4$ fb while we find the corresponding results $473^{+21}_{-30}$ fb and $235^{+10}_{-15}$ fb. Again, the results of the two calculations are in excellent agreement.

Next, we compare our results with data from the LHC at 8 and 13 TeV. We note that the data at both energies are consistently higher than the NLO cross sections.  At 8 TeV, CMS has measured a $t{\bar t}W$ production cross section of $382^{+117}_{-102}$ fb \cite{CMS8b} while ATLAS finds $369^{+100}_{-91}$ fb \cite{ATLAS8}. Both of these numbers are substantially above the NLO cross sections and even the aNNLO cross sections, and thus we need aN$^3$LO corrections for theory to describe the data. The aN$^3$LO QCD + NLO EW result is $266^{+7}_{-12}{}^{+6}_{-6}$ fb, where the first uncertainty is from scale variation and the second is from the pdf, and this is in good agreement with both the CMS and the ATLAS data.  

CMS has recently measured a $t{\bar t}W$ production cross section at 13 TeV of $868 \pm 65$ fb \cite{CMS13b}. Separate measurements were also made for $t{\bar t}W^+$ of $553 \pm 42$ fb and for $t{\bar t} W^-$ of $343 \pm 36$ fb \cite{CMS13b}. ATLAS has measured a cross section at 13 TeV of $880 \pm 80$ fb \cite{ATLAS13c}.  Separate measurements were also made for $t{\bar t}W^+$ of $583 \pm 58$ fb and for $t{\bar t} W^-$ of $296 \pm 40$ fb \cite{ATLAS13c}. All of these numbers are substantially above the NLO cross sections and even the aNNLO cross sections, and thus we again need aN$^3$LO corrections to reach agreement with theory. The aN$^3$LO QCD + NLO EW result is $779^{+22}_{-19}{}^{+12}_{-13}$ fb, where the first uncertainty is from scale variation and the second is from the pdf, and this is in good agreement with both the CMS and the ATLAS data. In addition, the aN$^3$LO QCD + NLO EW result for $t{\bar t}W^+$ is $517^{+14}_{-12}{}^{+8}_{-9}$ fb, in good agreement with both the CMS and ATLAS data, while for  $t{\bar t} W^-$ it is $262^{+8}_{-7}{}^{+4}_{-4}$ fb, in good agreement with the ATLAS data and bringing the theoretical prediction closer to the CMS data.

\mysection{Top-quark $p_T$ and rapidity distributions}

In this section we present results for the top-quark $p_T$ and rapidity distributions in $t{\bar t}W$ production. As for the total cross sections, we use a top-quark mass $m_t=172.5$ GeV and set the factorization and renormalization scales equal to each other, with this common scale denoted by $\mu$. Again, the complete NLO results include both QCD and EW corrections and are calculated using {\small \sc MadGraph5\_aMC@NLO} \cite{MG5,MGew}. We add second-order soft-gluon corrections to the NLO QCD and NLO QCD+EW results to derive aNNLO QCD and aNNLO QCD + NLO EW distributions, respectively; and we further add third-order soft-gluon corrections to the aNNLO QCD and aNNLO QCD + NLO EW results to derive aN$^3$LO QCD and aN$^3$LO QCD + NLO EW distributions, respectively. Again, we use MSHT20 NNLO pdf \cite{MSHT20} in our calculations.

\begin{figure}[htbp]
\begin{center}
\includegraphics[width=90mm]{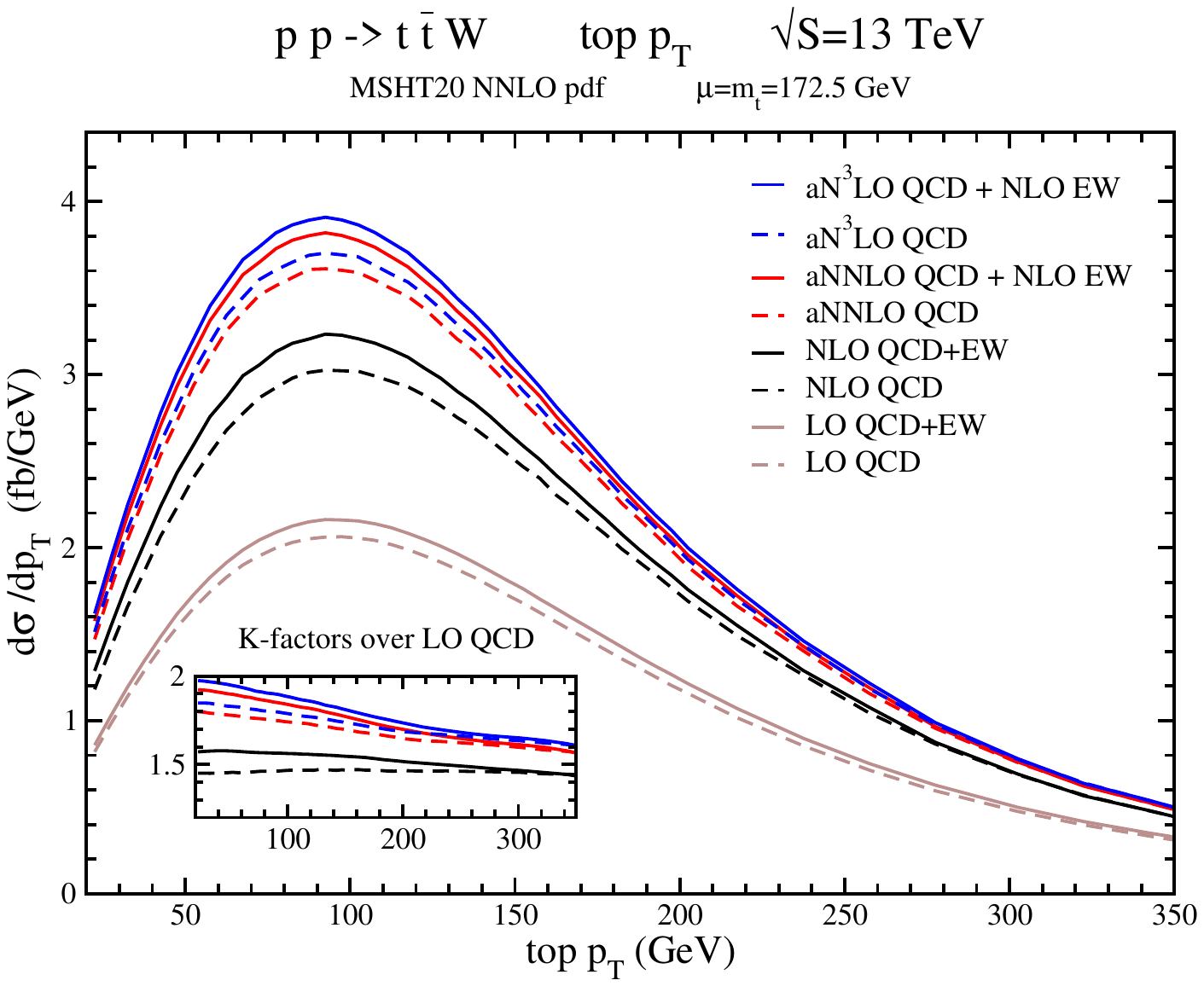}
\hspace{-5mm}
\includegraphics[width=90mm]{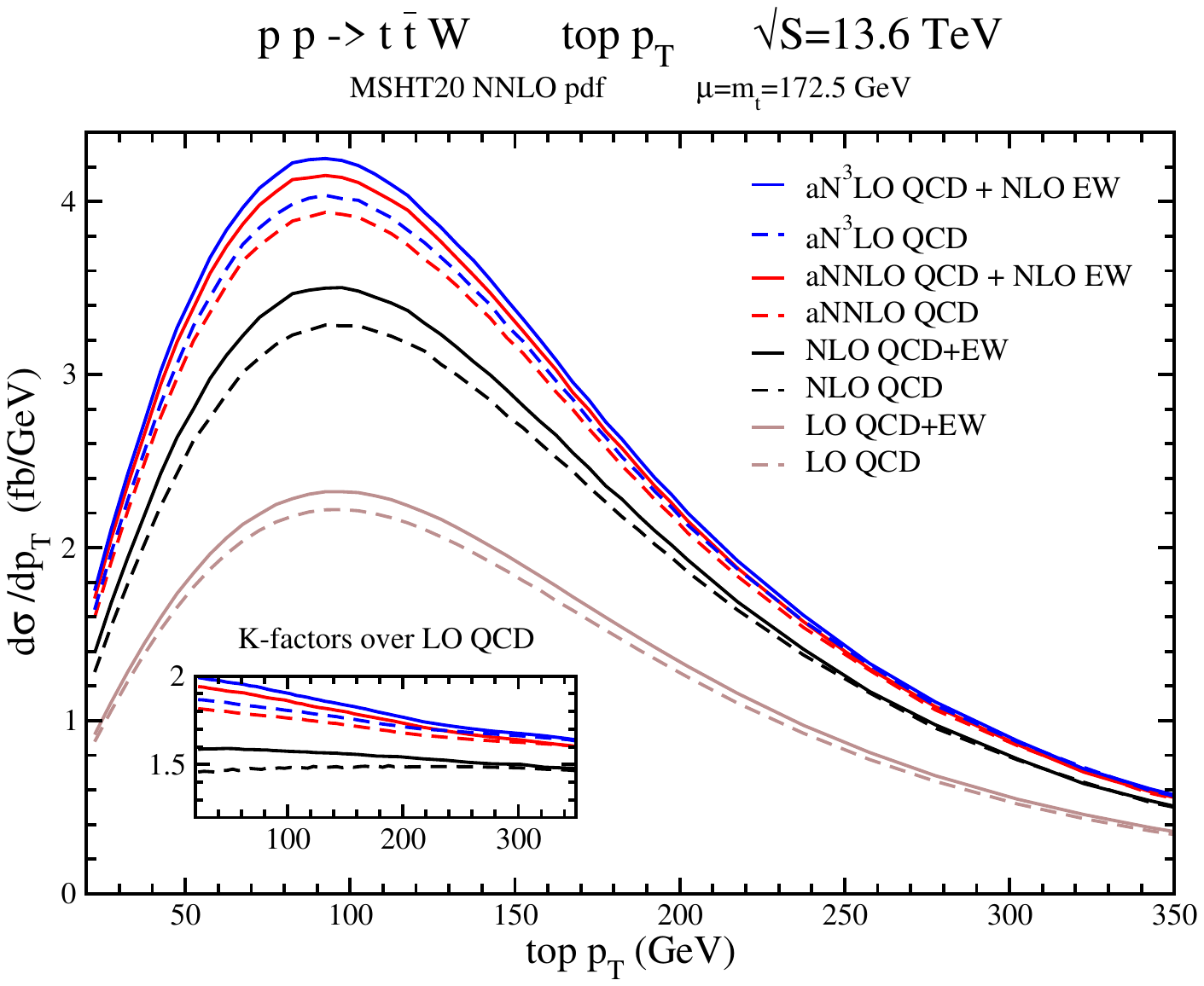}
\caption{The top-quark $p_T$ distributions at various orders through aN$^3$LO QCD + NLO EW in $t{\bar t}W$ production in $pp$ collisions at LHC energies of 13 TeV (left plot) and 13.6 TeV (right plot) using MSHT20 NNLO pdf. The inset plots display the $K$-factors relative to LO QCD.}
\label{pttop}
\end{center}
\end{figure}

In Fig. \ref{pttop} we plot the top-quark transverse-momentum distributions, $d\sigma/dp_T$, in $t{\bar t}W$ production at LHC energies of 13 TeV (left plot) and 13.6 TeV (right plot). Results are shown at various orders through aN$^3$LO QCD + NLO EW. The inset plot displays the $K$-factors, i.e. the ratios of the higher-order results to the LO QCD $p_T$ distribution. As for the total cross section, we see large enhancements in the $p_T$ distribution from the NLO QCD corrections and further significant increases from the higher-order soft-gluon corrections as well as the electroweak corrections. However, the relative sizes of these enhancements change with the value of the top-quark $p_T$.

We also note that the expansion of the resummed cross section to NLO provides very good approximation to the exact NLO result for the top-quark $p_T$ distribution. This is expected given the results for the total cross section and, also, given that the same is well known to hold for $t{\bar t}$ production. The difference between the two results is no more than a couple of percent in the $p_T$ range considered.

\begin{figure}[htbp]
\begin{center}
\includegraphics[width=90mm]{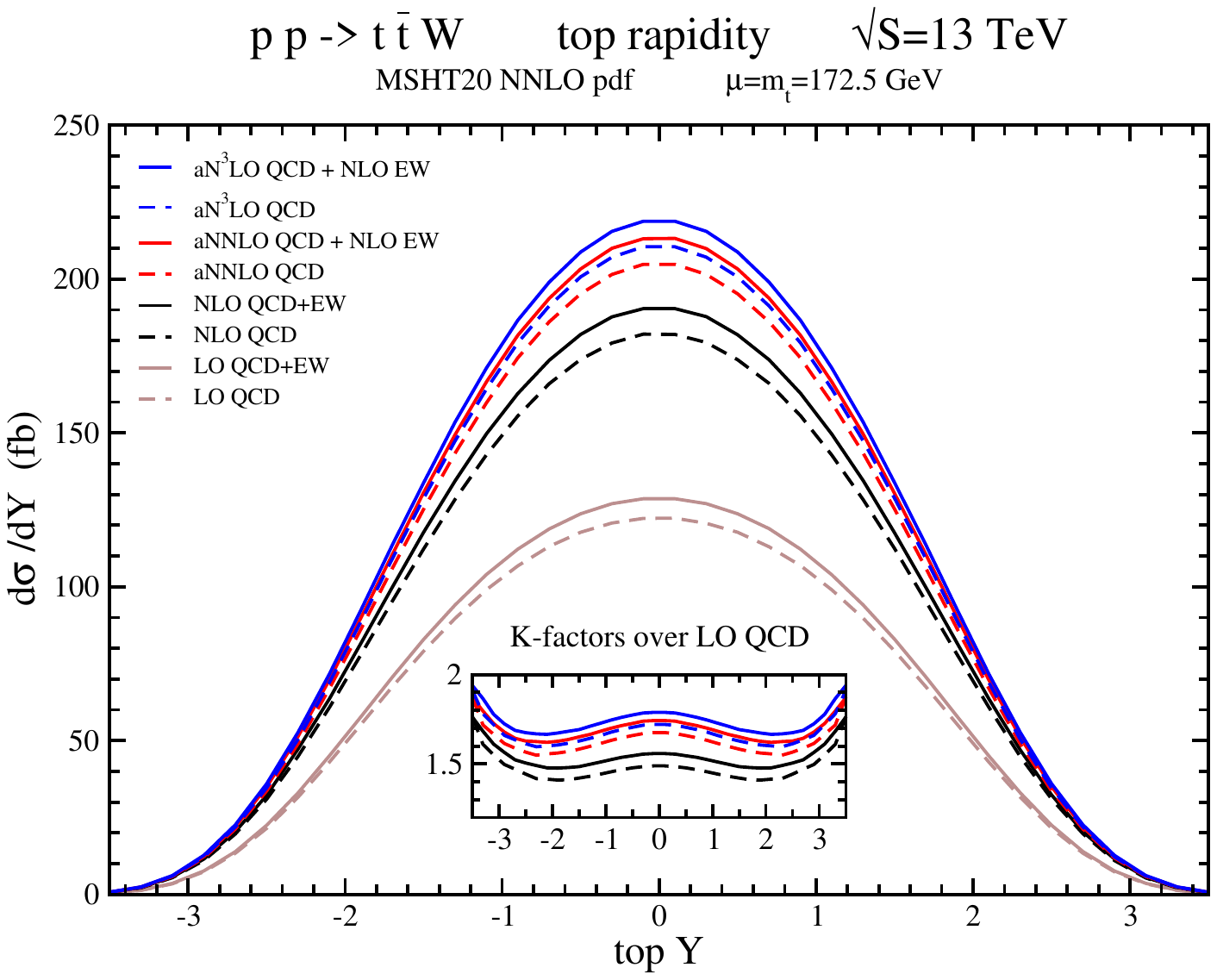}
\hspace{-5mm}
\includegraphics[width=90mm]{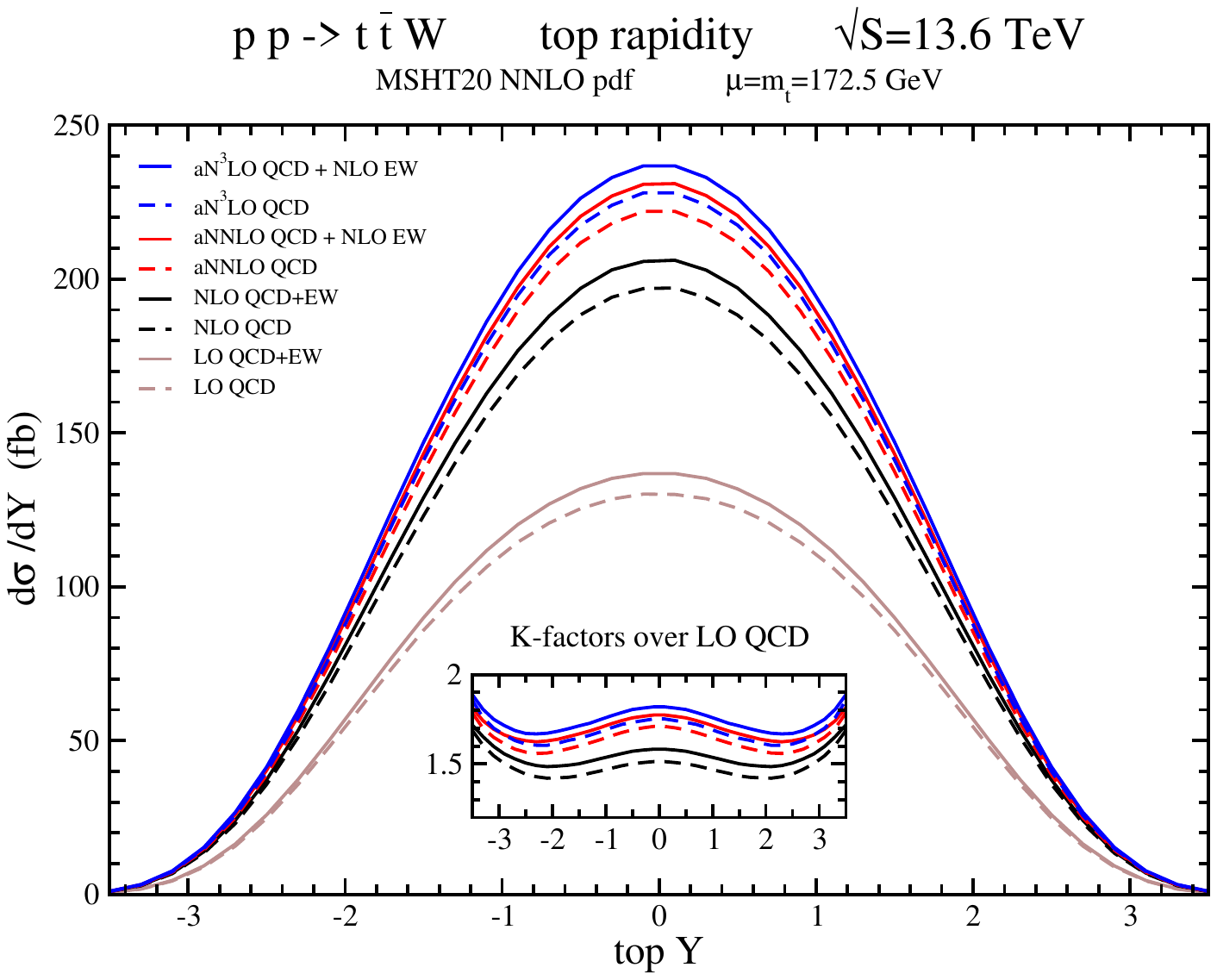}
\caption{The top-quark rapidity distributions at various orders through aN$^3$LO QCD + NLO EW in $t{\bar t}W$ production in $pp$ collisions at LHC energies of 13 TeV (left plot) and 13.6 TeV (right plot) using MSHT20 NNLO pdf. The inset plots display the $K$-factors relative to LO QCD.}
\label{ytop}
\end{center}
\end{figure}

In Fig. \ref{ytop} we plot the top-quark rapidity distributions, $d\sigma/dY$, in $t{\bar t}W$ production at LHC energies of 13 TeV (left plot) and 13.6 TeV (right plot). Results are shown at various orders through aN$^3$LO QCD + NLO EW. The inset plot displays the $K$-factors, i.e. the ratios of the higher-order results to the LO QCD rapidity distribution. Again, the contributions from the higher-order QCD corrections through aN$^3$LO are large, and the EW corrections are significant. The relative sizes of the enhancements from the higher-order corrections depend on the value of the top-quark rapidity and grow faster at very large rapidity values.

We also note that, similar to the case of the $p_T$ distribution, the expansion of the resummed cross section to NLO provides a very good approximation to the exact NLO result for the top-quark rapidity distribution. The difference between the approximate and exact NLO results is around two percent or less in the rapidity range considered.

\mysection{Conclusions}

We have provided a study of higher-order corrections for $t{\bar t} W$ production at the LHC. We have included exact QCD corrections at NLO as well as soft-gluon corrections at aNNLO and aN$^3$LO. We have also included electroweak corrections through NLO. We have calculated total cross sections as well as top-quark differential distributions for $t{\bar t}W$ production at LHC energies through aN$^3$LO QCD + NLO EW.

The total cross sections get large enhancements from the complete NLO QCD and electroweak corrections. The additional aNNLO and aN$^3$LO QCD corrections from soft-gluon emission are significant, accounting for further enhancements of the cross section. The theoretical uncertainties from scale variation get reduced with each increasing perturbative order. We have also provided results separately for $t{\bar t}W^+$ and $t{\bar t}W^-$ production. The aN$^3$LO corrections are necessary to find agreement with the LHC data at 8 and 13 TeV. 

We have also computed the top-quark $p_T$ and rapidity distributions for two recent center-of-mass LHC energies, namely 13 TeV and 13.6 TeV. These differential distributions also receive large QCD corrections at NLO and further significant enhancements at aNNLO and aN$^3$LO, as well as significant electroweak corrections. The effects of these enhancements depend on the value of the top-quark transverse momentum or rapidity.

\section*{Acknowledgements}
This material is based upon work supported by the National Science Foundation under Grant No. PHY 2112025.

\end{document}